\documentclass[aps,preprintnumbers,prl,twocolumn,superscriptaddress]{revtex4}

\usepackage{epsfig,latexsym,cancel,amssymb,amsmath}
\usepackage{graphicx}
\usepackage{epstopdf}
\usepackage{color}

\begin{document}
%\DeclareGraphicsExtensions{.jpg,.pdf,.mps,.png,.eps}

\preprint{CALT-TH-2015-028}

%\title{Searching for compressed top superpartners at 13 TeV LHC}
\title{Opening up the compressed region of stop searches at 13 TeV LHC }

 \author{Haipeng An}
%\email[e-mail: ]{}
\affiliation{Walter Burke Institute for Theoretical Physics, California Institute of Technology, Pasadena, CA 91125}

\author{Lian-Tao Wang}
%\email[e-mail: ]{}
\affiliation{Enrico Fermi Institute and Department of Physics
University of Chicago, 5620 S. Ellis Avenue, Chicago, IL 60637, USA}
\affiliation{Kavli Institute for Cosmological Physics
University of Chicago, 933 E. 56th Street, Chicago, IL 60637, USA}

%\today

\begin{abstract}
Light top superpartners play a key role in stabilizing the electroweak scale in supersymmetric theories. For $R$-parity conserved supersymmetric models, traditional searches are not sensitive to the compressed regions. In this paper, we propose a new method targeting this region, with stop and neutralino mass splitting ranging from  $m_{\tilde t} -  m_\chi  \gtrsim  m_t$ to  about 20 GeV. 
In particular, we focus on the signal process in which a pair of stops are produced in association with a hard jet, and define a new observable $R_M$ whose distribution has a peak in this compressed region. The position of the peak is closely correlated with $m_{\tilde t}$. We show that for 13 TeV LHC with a luminosity of 3000 fb$^{-1}$, this analysis can extend the reach of stop in the compressed region to $m_{\tilde t}$  around 800 GeV. 
\end{abstract}

\maketitle

\paragraph{Introduction}
\label{sec:introduction}

With the discovery of the Higgs~\cite{Chatrchyan:2012ufa,Aad:2012tfa}, particle physics reached an important milestone. However, the mechanism of stabilizing the electroweak scale from large quantum corrections is one of the outstanding mysteries. 
In most of the models addressing this problem, a key ingredient is a light top partner. 
As the most prominent example,  in supersymmetry, the stop $\tilde t$ should be less than about a TeV to be an effective solution to the fine-tuning problem \cite{Papucci:2011wy}.
Traditional searches for stops focus on pair production of stops with each of them decays into $t$ and the LSP, $\chi$. If $m_{\tilde t} \gg m_\chi + m_t$,   the top quark can be quite energetic. 
In the top pair production, a main background of this search,  most of the top quarks are produced near the threshold.  
 Based on this observation, various kinematical variables (e.g. $m_{T2}$~\cite{Lester:1999tx,Barr:2003rg,Bai:2012gs,Kilic:2012kw,Cao:2012rz}, $H_T$~\cite{Chatrchyan:2013wxa}, the razor variables~\cite{Rogan:2010kb,Chatrchyan:2012uea,Khachatryan:2015pwa} and the variables invented in Ref.~\cite{Nachman:2013bia}) have been defined to distinguish stop pair production from top pair production. For recent global studies of the minimal version of the supersymmetric standard model, see \cite{deVries:2015hva} and references therein. However, in the compressed region where $m_{\tilde t} \approx m_t + m_\chi$, the kinematics of the top quarks from stop decay are similar to those in the top pair production, and such observables are less sensitive. In the region that $m_\chi \ll m_{\tilde t} \approx m_t$, spin correlations of the top quarks  can help to distinguish the signal from background~\cite{Jezabek:1994qs,Brandenburg:2002xr,Han:2012fw}. Such analysis has been done by the CDF, D0, ATLAS and CMS collaborations~\cite{Aaltonen:2010nz,Abazov:2011qu,Abazov:2011ka,Abazov:2011mi,ATLAS:2012ao,Aad:2014pwa,Chatrchyan:2013wua,Aad:2014mfk}. However, with larger $m_{\tilde t}$ this method does not work well due to smaller production rate. In another extreme regions of the parameter space $m_{\tilde t} \approx m_\chi$, $\tilde t$  decays into 4 body final states or a light quark plus the LSP through flavor-changing processes.
In the case that the flavor-changing processes are important, charm tagging can be useful~\cite{Choudhury:2012kn,Belanger:2013oka,Aad:2014nra}.
However, the jets from the decay are usually soft and cannot be identified. The leading search channel is mono-jet + MET~\cite{Carena:2008mj,Bornhauser:2010mw,Ajaib:2011hs,Drees:2012dd,Dreiner:2012sh,Krizka:2012ah,Delgado:2012eu,Cohen:2013zla,Low:2014cba,Ferretti:2015dea,Khachatryan:2015wza,Hikasa:2015lma}.  Light stops can also be probed directly by comparing the observed $t\bar t$ or $W^+W^-$ pair production rate with theoretical calculations~\cite{Czakon:2014fka,Aad:2014kva,Rolbiecki:2015lsa}. However, it will be difficult for this method to be benefited from larger luminosity and higher energies in the future runs of LHC, since its sensitivity is mainly limited by systematic errors. Vector boson fusion tagging has also been proposed to search for stop in the compressed region, and it has been shown that it is still cannot fully close the gap in the compressed region~\cite{Dutta:2013gga}.
If the life-time of the stop is long enough, 
a pair of stops can form a bound state, the stoponium. In this case, searches of the stoponium can be sensitive to these compressed regions~\cite{Drees:1993yr,Drees:1993uw,Martin:2008sv}. For recent detailed studies of LHC sensitivities see~\cite{Batell:2015zla} and references therein. 
In this region, the stops can also hadronize first and then decay with displaced vertices~\cite{Grober:2014aha}.
If the heavier stop is reachable, one can also study the decay of the heavier stop to the Higgs boson or $Z_0$ together with the lighter stop~\cite{Perelstein:2007nx,Ghosh:2013qga,Khachatryan:2014doa,Aad:2015pfx}. 
{If the sleptons or charginos are lighter than the compressed stop, the decay pattern of the stop can be changed dramatically. See Ref.~\cite{Padley:2015uma} for recent study of these scenarios. }
Constraints on masses of the two stops can also be inferred from the measurement of the Higgs mass and  production rate~\cite{Han:2013kga,Fan:2014txa}. 
Light stops also get constraints from low energy precision experiments such as the $b\rightarrow s\gamma$ experiment~\cite{Degrassi:2000qf,Ishiwata:2011ab,Blum:2012ii}.  

However, it is still difficult for current searches to cover the compressed region with mass splitting ranging from  $m_{\tilde t} - m_\chi \gtrsim m_t $ to much smaller values about 20 GeV. 
In this letter, we introduce a new kinematical observable which targets the kinematics of this compressed region, and demonstrate its effectiveness. We note that there are other studies focusing on the similar parameter region~\cite{Alves:2012ft,Hagiwara:2013tva,Delgado:2012eu}. In particular, the strategy adopted in Ref.~\cite{Delgado:2012eu} can cover parameter space around $m_{\tilde t} \approx m_W + m_b + m_\chi$, although it is less effective for $m_{\tilde t} \approx m_t + m_\chi$.
%\haipeng{We need to rewrite this paragraph to incorporate Ref [36] (Delgado et al), Ref [60] (the Japanese paper) Ref [61] Fox and Buckley}
% \liantao{ref wording added.}
 
\paragraph{Kinematics around the compressed region}
\label{sec:kinematics}
\bigskip

In the compressed region $m_{\tilde t} \gtrsim m_t + m_\chi$, the $\tilde t$ first decays into a pair of $t$ and $\chi$. Due to the compressed nature, in the rest frame of $\tilde t$, the $\chi$ and $t$ are almost at rest. Therefore, in the lab frame,  the transverse momenta of the $\tilde t$ and $\chi$ have a simple relation that $\vec p_T(\chi) \simeq (m_\chi / m_{\tilde t}) \vec p_T(\tilde t)$. Therefore, in the process of the $\tilde t$ pair production, 
the contribution to MET from the two $\chi$'s approximately cancel each other, and as a result the kinematics of the top quarks from $\tilde t$ pair production is very similar to those from the top pair production, making the search very difficult. We propose to focus on events with an additional hard jet from initial state radiation (ISR). In this case, we have $\vec p_T(j_{\rm ISR}) \simeq - (\vec p_T(\tilde t_1) + \vec p_T(\tilde t_2))$ where $\tilde t_1$ and $\tilde t_2$ are the two stops produced in this process. Therefore, the ratio between MET and the $p_T(j_{\rm ISR})$ 
\begin{equation}\label{eq:RM}
R_M \equiv \not\! p_T / p_T(j_{\rm ISR}) \approx {m_\chi}/{m_{\tilde t}}  \ ,
\end{equation}
where $\not\!\! p_T$ is the total MET in this process. Hence, we expect a peak-like feature in the $R_M$ distribution. {Ref.~\cite{Hagiwara:2013tva} has also noticed  the similar kinematical feature.}
The spread of this peak can come from several sources. 
In the rest frame of $\tilde t$, the momentum acquired by $\chi$ can be written as
\begin{equation}
\Delta p_\chi = \frac{[(m_{\tilde t}^2 - (m_t + m_\chi)^2) (m_{\tilde t}^2 - (m_t - m_\chi)^2) ]^{1/2}}{2 m_{\tilde t}} \ ,
\end{equation}
Therefore, in the compressed region,
\begin{equation}\label{eq:deltap}
\Delta p_\chi \approx \left(\frac{2 m_t m_\chi \Delta m}{m_{\tilde t}}\right)^{1/2} \lesssim (2m_t \Delta m)^{1/2} \ ,
\end{equation}
where $\Delta m \equiv m_{\tilde t} - m_\chi - m_t$. In most part of the parameter space we are interested in in this study, the boosts of the $\tilde t$'s are small. Therefore
\begin{equation}\label{eq:parton}
\Delta R_{\rm parton} \equiv \Delta p_\chi / p_T(j_{\rm ISR})
\end{equation}
can serve as a good estimate of the width of peak in the $R_M$ distribution at parton level. 
In practice, additional soft radiation and detector effects will also smear the distribution of $R_M$. 
Nevertheless, as we will demonstrate, there is still a peak  in the $R_M$ distribution around $m_\chi / m_{\tilde t}$ in the compressed region. 

In the compressed region where $m_{\tilde t} \lesssim m_t + m_\chi$, the $\tilde t$ decays into $\chi$ and $b$ and $W$ through a virtual $t$. Neglecting the spin correlation between the initial and final states the differential decay width of $\tilde t$ with respect to the invariant mass of the virtual $t$ can be written as
\begin{equation}\label{eq:dis}
\frac{d\Gamma_{\tilde t}}{d q_t} \approx \frac{\Gamma^{(2)}_{\tilde t}(q_t)}{\pi} \frac{{q_t}^2 \Gamma_t(q_t)}{(q_t^2 - m_t^2)^2} \ ,
\end{equation}
where $q_t$ is the virtual mass of the top quark. $\Gamma^{(2)}_{\tilde t}(q_t)$ is the two-body decay width of $\tilde t$ with replacement $m_t \to q_t$, and $\Gamma_t(q_t)$ is the decay width of $t$, replacing $m_t \to q_t$.  Eq.~(\ref{eq:dis}) implies that $q_t$ prefers to be as close to $m_t$ as possible, with maximal value $q_{t \rm max}=m_{\tilde t} - m_{\chi}$. Hence, even the top quarks are virtual,  the $\chi$ decayed from each $\tilde t$ still prefers to be at rest in the rest frame of $\tilde t$. Therefore, the relation shown in Eq.~(\ref{eq:RM}) still holds approximately. 
Since we are not far away from the region where the top is on-shell, we expect the spread of the peak at parton level is still around the value given by Eq.~(\ref{eq:deltap}), with replacement $\Delta m \to |\Delta m|$. 

Similarly, we also expect to see a sharp peak in the $R_M$ distribution at the compressed region $m_{\tilde t} \approx m_W + m_b + m_\chi$, where the W boson and the LSP is approximately stationary in the stop rest frame. The width of the peak  in the $R_M$ distribution at parton level can be estimated using Eq.~(\ref{eq:parton}) with $m_t$ in Eq.~(\ref{eq:deltap}) replaced by $m_W$. 

\paragraph{SM background and basic cuts}
\label{sec:numerical}
\bigskip

For this analysis, it is crucial to identify which jet is from ISR. $\tilde t$s with a mass of several hundred GeV are not usually highly boosted. Therefore, in the compressed region, the $t$s from the $\tilde t$ decay are also not highly boosted. Thus, the $p_T$ of the hardest jet in the decay chain of the $\tilde t$ is around $m_t$. As a consequence, if we require  $p_T(j_0)\gg m_t$, where $j_0$ is the hardest jet,
we found that it is very probable that $j_0$ is $j_{\rm ISR}$. Hence, we will use the ratio $\not\! p_T /p_T(j_0)$ as an approximation for $R_M$.  The requirement of a large $p_T(j_0)$ also helps reduce the QCD background and sharpen the peak of the $R_M$ distribution as shown in Eq.~(\ref{eq:parton}). In practice, we require $p_T(j_0) > 700$ GeV. 

The leptonic decay of $t$ is always accompanied with neutrinos, which smears the peak structure in the $R_M$ distribution. In this analysis, we focus on the hadronic decays and veto events with charged leptons. At parton level, 6 soft jets from top decay appears. In practice, some of these soft jets may merge into a harder one. Therefore, we require at least three sub-leading jets with $p_T > 60$ GeV. 

An important kinematical feature of the signal is that, in the compressed region, the $\vec p_T(j_{\rm ISR})$ is approximately in the opposite direction as the $\not\!\!\vec p_T$. Therefore, we  require $|\phi(j_0) - \phi_{\rm MET} - \pi| < 0.15$.  At the same time, we expect a significant  QCD background from the mis-measurement of jet energy.  To  reduce the background due to the mis-measurement of the subleading jets,  we require $|\phi_{\rm MET} - \phi_j| > 0.2$ for all the jets with $p_T > 60$ GeV. Requiring $p_T > 60$ GeV also helps to reduce the pile-up effects which are significant during the high luminosity runs of the LHC.

\begin{figure}[tb]
\centering
\includegraphics[width=\columnwidth]{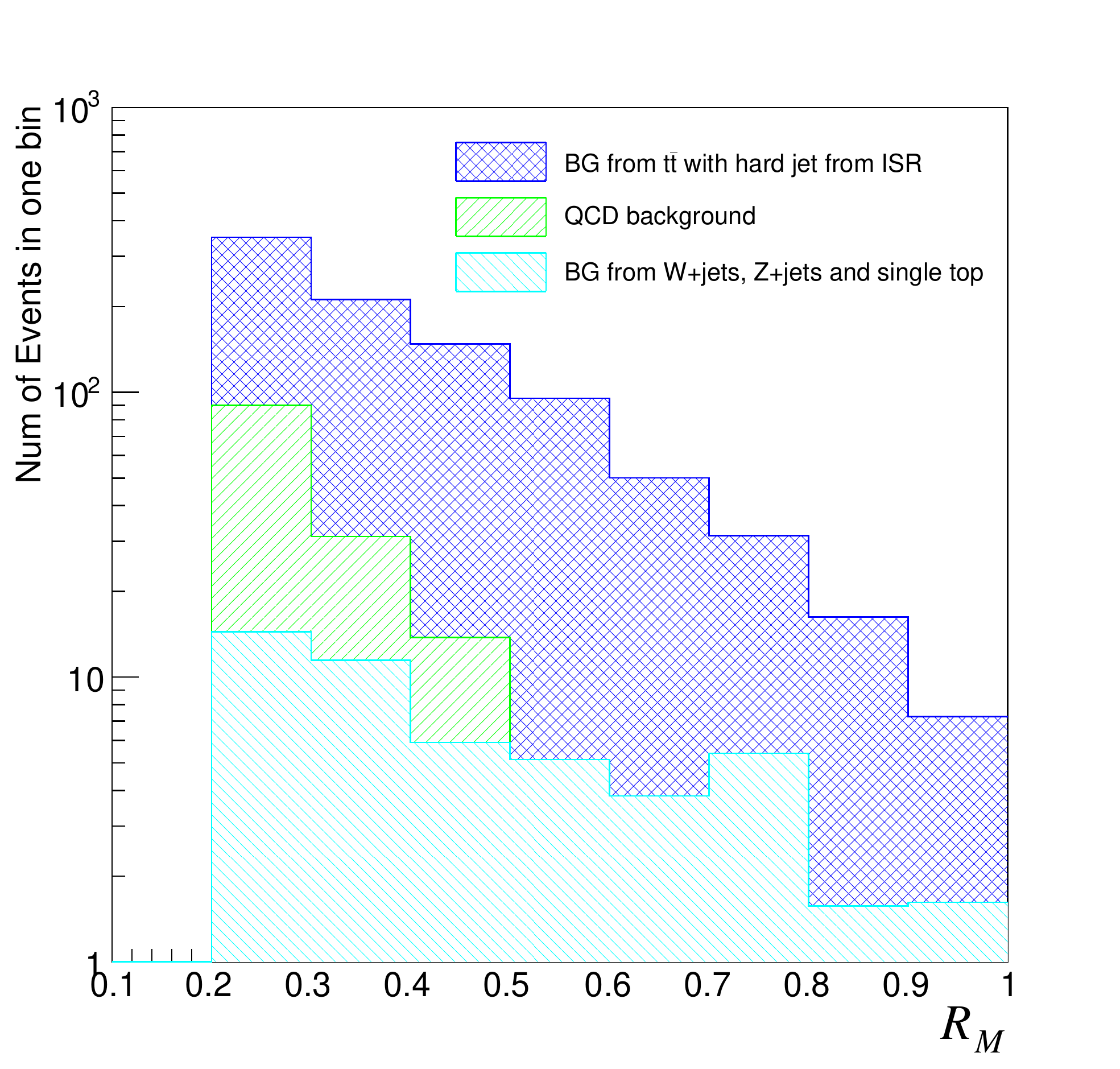}
\caption{Contributions to background from various processes,  after the basic cuts described in the text. }
\label{fig:bg}
\end{figure} 

\begin{figure}[tb]
\centering
\includegraphics[width=\columnwidth]{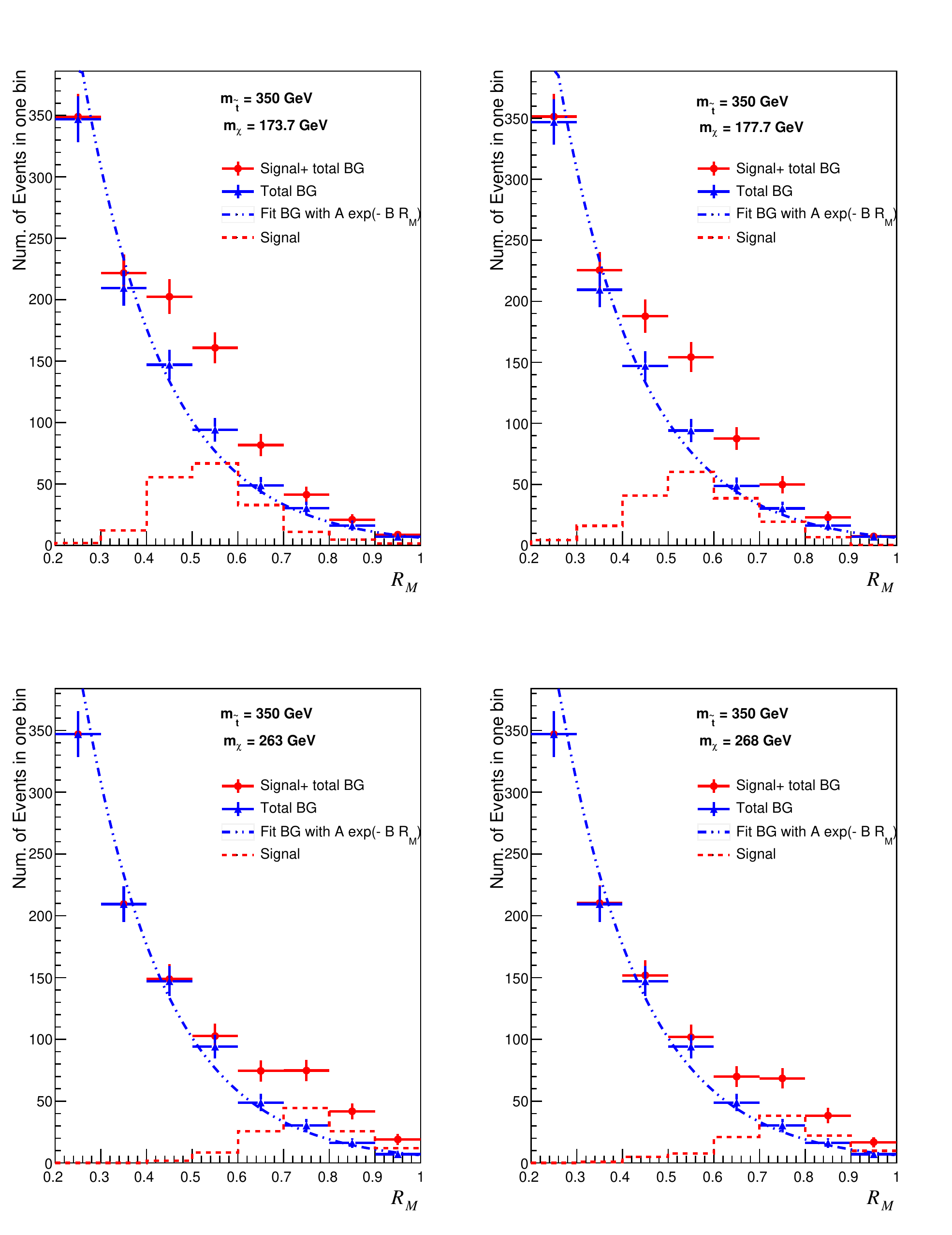}
\caption{$R_M$ distributions for $m_{\tilde t} = 350$ GeV on both sides of the $m_{\tilde t} = m_{\chi} + m_t$ (top) and $m_{\tilde t} = m_\chi + m_W + m_b$ (bottom). 
}
\label{fig:RM}
\end{figure}

\begin{figure*}[tp]
\centering
\includegraphics[width=\columnwidth]{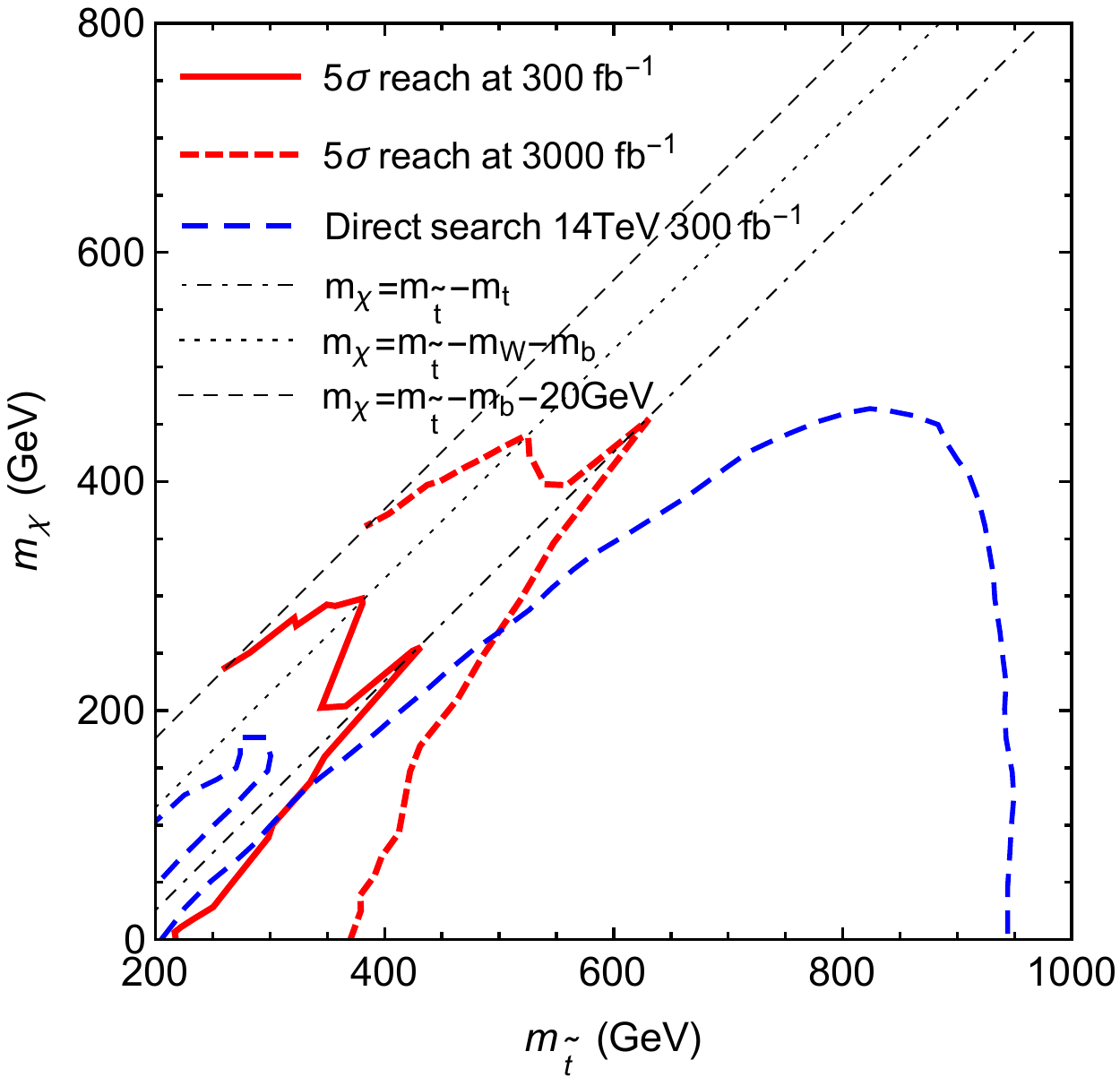}
\includegraphics[width=\columnwidth]{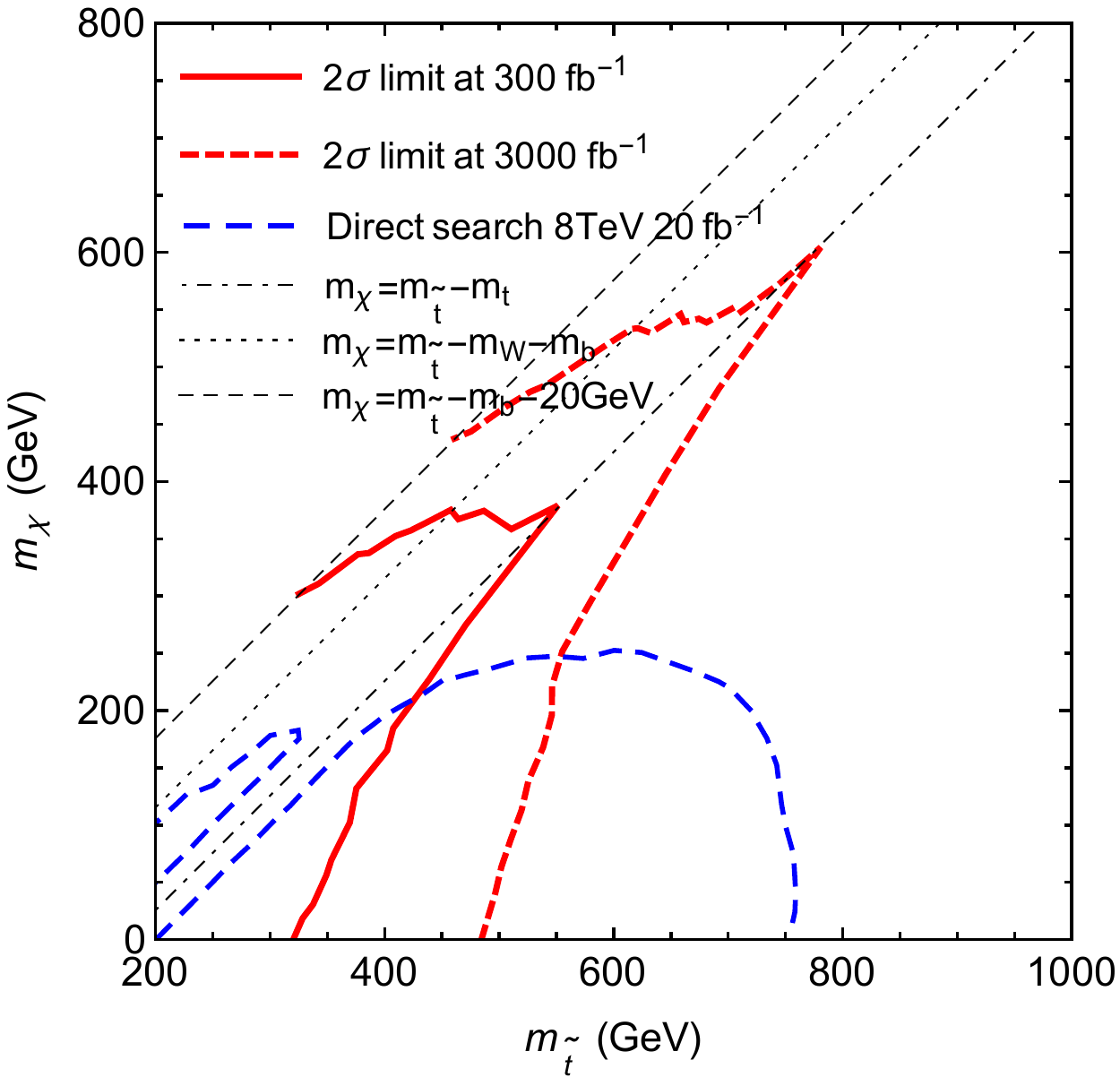}
\caption{The projected 5 $\sigma$ discovery  reach (left) and 95\% C.L. exclusion limit (right) of stop in the compressed region. }
\label{fig:bounds}
\end{figure*}

To further reduce the QCD background, we require at least one b-jet appears in the final state. The b-tagging efficiency we use is the {\it tight} b-tagging in the PGS detector simulator~\cite{PGS}, in which the b-tagging efficiency is about $40\%$ for $p_T(b)$ around 100 GeV and within $|\eta| < 1.2$. 
For such a tagging efficiency the mistag rate of light partons 
for CMS detector can be as small as $0.1\%$~\cite{cms_btag}. Detailed simulation shows that the main QCD background is from the processes with $b$ and $c$ quarks in the final states. Since it is very unlikely for $b$ jet to be the leading ISR jet in the signal, we veto events with the leading jet passed the {\it loose} b-tagging in the PGS detector simulation~\cite{PGS}, which is around 45\% for jet with $p_T > 700$ GeV. This is smaller than the current benchmark b-tagging efficiency of the CMS and ATLAS detectors, therefore our result is conservative.

\paragraph{Numerical results}
\label{sec:numerical}
\bigskip
For both the background and signal, the parton level simulations are done using MadGraph5/MadEvent~\cite{Alwall:2011uj} followed by parton shower with PYTHIA6.4 \cite{Sjostrand:2006za}. The detector simulation is done using PGS4 \cite{PGS} with anti-$k_T$ jet algorithm with a distance parameter of 0.5~\cite{Cacciari:2008gp}. For the background, the MLM matching scheme is also used to avoid double counting~\cite{Mangano:2002ea}. For signal we checked the results from simulations with and without matching that the difference is within 20\%.
With all the basic cuts discussed above, the $R_M$ distribution from SM processes with the cuts previously described is shown in Fig.~\ref{fig:bg}. The dominant contribution to the background is from $t\bar t$ pair production with a hard ISR jet. In our signal region with large $R_M$, a significant amount of MET is required. Since we veto events with charged leptons in the final state, the dominant contribution to the background is from leptonic decays of top with $\tau$s, or mis-tagged $e/\mu$s.
The second leading background comes from QCD multi-jet production with at least one of the jets containing a bottom or charm quark. The background from electroweak processes is not important due to their smaller rates.

From Fig.~\ref{fig:bg} one can see that both the $t\bar t+{j_{\rm  ISR}}$ background and the QCD background exponentially decrease with $R_M$ due to the lack of the source of MET. The background from electroweak processes is relatively flat but with a suppressed rate as shown in Fig.~\ref{fig:bg}. The total background is well fitted by a function 
\begin{equation}
\frac{d\sigma}{d R_M} = A \exp (-B R_M) \ ,
\end{equation}
where $A$ and $B$ depends on the details of the cuts, and in the current choice $A=47$ fb and $B=5.6$.

For the signal, the $R_M$ distribution for $m_{\tilde t} = 350$ GeV and several different $m_{\chi}$
are shown in Fig.~\ref{fig:RM}. To make the feature easier to visualize, we choose points  very close to the mass thresholds with $\Delta m \approx 2$ GeV.  One can see that in all cases the $R_M$ distribution is peaked at around $m_{\chi}/m_{\tilde t}$, with widths around 0.2. From Eq.~(\ref{eq:dis}),  the width generated by the phase space of the decay of $\tilde t$ is about 0.05. Therefore,  the typical width of the peak of the $R_M$ distribution induced by parton shower and detector effect is about $0.2$.

In order to take advantage of the peak in the $R_M$ distribution, and the fact that the background decays exponentially with $R_M$ we add another cut that for $m_{\tilde t} < m_{t} + m_\chi$
\begin{equation}\label{eq7}
\left(\frac{m_\chi}{m_{\tilde t}}\right) - 0.05 < \frac{\not\! p_T}{p_T(j_0)} < \left(\frac{m_\chi}{m_{\tilde t}}\right) +0.15 \ ,
\end{equation}
and for $m_{\tilde t} > m_t + m_\chi$
\begin{equation}\label{eq8}
\left(\frac{m_{\tilde t} - m_t}{m_{\tilde t}}\right) - 0.05 < \frac{\not\! p_T}{p_T(j_0)} < \left(\frac{m_{\tilde t} - m_t}{m_{\tilde t}}\right) +0.15 \ .
\end{equation}
As the $m_\chi\rightarrow0$,  the background in the cut window defined in Eq.~(\ref{eq7}) grows exponentially. Therefore, for the region $m_{\tilde t} > m_t + m_\chi,$ we choose a different window as shown in Eq.~(\ref{eq8}) which is independent of $m_\chi$, so that more parameter space in the bulk region can be covered effectively.
We define the $5\sigma$ and $2\sigma$ expected limit by $S/\sqrt{B} = 5$ and $S/\sqrt{B+S} = 2$, where $S$ and $B$ are the signal and background after all the cuts. The $5\sigma$ reach at CMS with the center-of-mass energy of 13 TeV and luminosities of 300 and 3000 fb$^{-1}$ are shown in left panel of Fig.~\ref{fig:bounds} together with the expected $5\sigma$ sensitivity of the direct stop pair production at 14 TeV LHC with CMS detector~\cite{CMS:2013xfa}. The right panel shows the $2\sigma$ expected 95\% C.L. exclusion limit together with the current combined  limit from direct stop pair production~\cite{Khachatryan:2015wza,Chatrchyan:2013xna,cms_sus14011,cms_sus13015}. The current limits and prospective reach of ATLAS are similar to CMS~\cite{Aad:2014bva,Aad:2014kra,Aad:2014qaa,ATLAS:2013hta, Aad:2014nra}, which are not shown here.
One can see that $\tilde t$ in the compressed region with a mass around 600 GeV can  be discovered in the LHC with 3000 fb$^{-1}$. 
It can also exclude stop with mass up to about 800 GeV.  Notice that there are "spike-like" features around the thresholds, $m_{\tilde t} - m_\chi \approx m_t$ and $m_{\tilde t} - m_\chi \approx m_W + m_b$, due to the fact that the peak in  $R_M$ is sharper around these thresholds.
\bigskip
\paragraph{Conclusion and discussion}

We point out a useful kinematical feature in the production of stop associate with an ISR jet, which can enhance the sensitivity in the compressed regions, with mass splitting ranging from  $m_{\tilde t} - m_\chi \approx  m_t$ to about 20 GeV.
We show that in this region the observable $R_M$ defined in Eq.~(\ref{eq:RM}) has a peak around $m_\chi/m_{\tilde t}$. Using this kinematical feature, we estimated that this gap can be covered up to around 800 GeV with 13 TeV LHC at a luminosity of 3000 fb$^{-1}$.

Although we have focused on the stop searches, the same technique is obviously applicable to the search of other top partner signals with similar final states. 

In the discussion, we neglect the flavor changing decay mode $\tilde t \rightarrow \chi c$, since it is model dependent and strongly constrained by flavor physics. It has been shown that in the minimal flavor violation scenario the branching ratio of this process is often subdominant to the four body decay of the stop in the region that $m_{\tilde t} < m_\chi + m_W + m_b$~\cite{Krizka:2012ah}. For detailed next-to-leading-order studies of the stop decay pattern in this region, see \cite{Muhlleitner:2011ww,Grober:2014aha,Grober:2015fia}.
This method is also applicable to other decay chains of stop in the compressed region~\cite{Bai:2013ema}.

The main background of this analysis is from top pair production associated with ISR jets, with at least one of the top decays leptonically and the charged lepton fails the lepton veto. A majority of these events  have the charged lepton close to hadronic activities. One may be able to distinguish these events further with alternative lepton isolation criteria. 
The top quarks generated from the stop decay in the signal are in general with a smaller boost than the top quarks from the background since $m_{\tilde t} > m_t$, which is assumed in this analysis. We may be able to use this property to further distinguish the top quarks from the signal and from the background. We leave these analysis to future work.

\bigskip
\paragraph{Acknowledgements} We would like to thank Clifford Cheung, Lance Dixon, David Kosower, Matthew Low, Yue Zhang and Fei Gao for useful discussions. H.A. is supported by the Walter Burke Institute at Caltech
and by DOE Grant DE-SC0011632. L.-T.W. is supported by DOE grant DE-SC0003930.

\end{document}